\documentclass[aps,pra,a4paper,showpacs,twocolumn]{revtex4}

\usepackage[latin1]{inputenc}
\usepackage{amsmath}
\usepackage{amsfonts}
\usepackage{amssymb}
\usepackage{bm}
\usepackage{graphicx}
\usepackage{epsfig}
\usepackage{subfigure}
\usepackage{setspace}
\usepackage{ulem}

\begin{document}
\renewcommand{\>}{\rangle}
\newcommand{\<}{\langle }
\title{Decoherence-free manipulation of photonic memories for quantum computation}
\date{\today}
\pacs{03.67.Lx, 42.50.Gy, 32.80.Qk}
\author{N. Sangouard}
\email{sangouard@physik.uni-kl.de}
\affiliation{Fachbereich Physik, Universt\"at Kaiserslautern, Erwin-Schroedinger Strasse, D-67663 Kaiserslautern, Germany}

\begin{abstract}
We present a protocol to construct an arbitrary quantum circuit. The quantum bits (qubits) are encoded in polarisation states of single photons. They are stored in spatially separated dense media deposed in an optical cavity. Specific sequences of pulses address individually the storage media to encode the qubits and to implement a universal set of gates. The proposed protocol is decoherence-free in the sense that spontaneous emission and cavity damping are avoided. We discuss a coupling scheme for experimental implementation in Neon atoms. 
\end{abstract}

\maketitle
\section{Introduction}
Quantum computation requires devices adapted to the transport, the memorisation and the manipulation of quantum information. Two circular polarization states of a single photon can be used to encode and to carry the information. With electromagnetically induced transparency (EIT)-type techniques \cite{lukin_revmodphys}, the photonic qubits then can be trapped, stored and released. Linear optical elements allow one to build a universal set of non-deterministic photonic gates \cite{knill_nature01}. However, the use of two specific devices to store and to manipulate photonic qubits constitutes a huge technological challenge. We thus wish \textit{to dispose of a unique system for both memorisation and manipulation of photonic qubits}. In this context, a two-qubit controlled phase gate ($\hbox{Cphase}$ gate) has been proposed \cite{lukin_PRL00_Masalas_PRA04} by using the strong coupling of two light pulses propagating in a storage medium. In this paper, we present a complete protocol for quantum computation allowing: (i) \textit{The storage} of the photonic qubits in dense atomic media. (ii) \textit{The qubit encoding}, i.e. the preparation of the qubit set in any superposition states. (iii) \textit{The computation} by implementing a universal set of gates from which an arbitrary quantum circuit can be constructed. Each process based on adiabatic passage, is \textit{robust} with respect to variations of several pulse parameters such as pulse area, duration etc \cite{vitanov_adv01}. Since the dynamics follows only dark states, i.e. states without any components on the excited states, the proposed protocol is \textit{decoherence-free} with respect to radiation decay.\\
We consider a quantum register composed of M single-photon pulses (see \cite{cirac_prl97} for the generation of single-photon pulses). Each photon has a well defined elliptic polarization described by the generic state $\alpha |1_{\hbox{-}},0_{+}\rangle+\beta |0_{\hbox{-}},1_{+}\rangle,$
$\alpha$ and $\beta$ being complex numbers such that $|\alpha|^2+|\beta|^2=1.$ $|1_{\pm}\rangle$ $(|0_{\pm}\rangle)$ is a Fock state associated to one (zero) photon of circular polarization $\sigma_{\pm}.$ The states $\{ |1_{\hbox{-}},0_{+}\rangle, |0_{\hbox{-}},1_{+}\rangle\}$ constituting to the computational basis are arbitrarily associated with the boolean states $\{|0\rangle, |1\rangle\}$ of a qubit. Each of the M photons interacts with a medium of N atoms initially prepared in the ground state $|b\rangle$ (Fig. \ref{system_universel}). The M media are placed in an optical cavity and can be addressed individually by a set of laser pulses. We first present the storage process and the encoding mechanism. We then give the sequence of pulses offering the possibility to implement a universal set of gates. We use a unit system where $\hbar=1.$\\
\section{Storage and Encoding}
\begin{figure}[ht!]
{\includegraphics[scale=0.45]{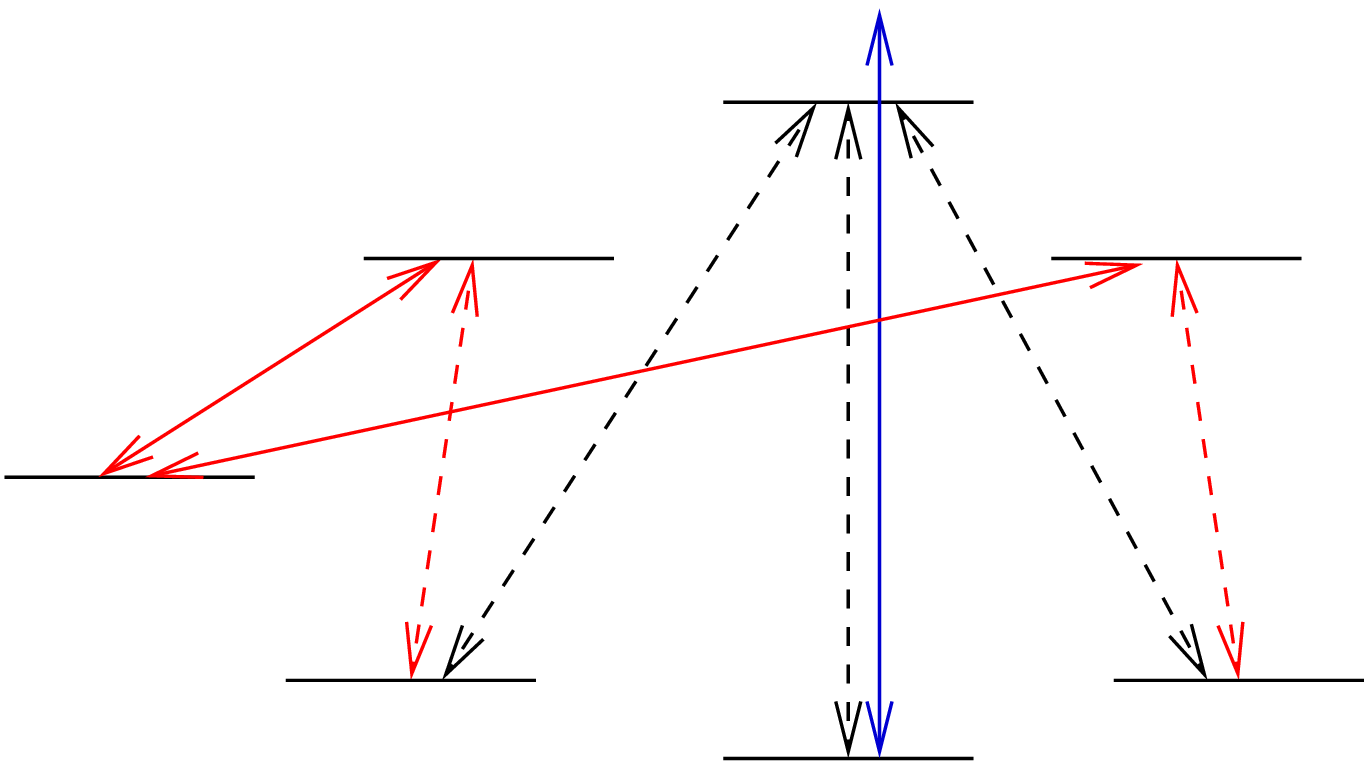}} 
\put(-167,25.5){$|b\rangle$}
\put(-122.5,70.5){$|a_{\hbox{-}}\rangle$}
\put(-37.5,70.5){$|a_{+}\rangle$}
\put(-80.5,93.5){$|e\rangle$}
\put(-74.5,-8.5){$|d\rangle$}
\put(-130.5,1.5){$|c_{\hbox{-}}\rangle$}
\put(-27.5,1.5){$|c_+\rangle$}
\put(-152,53.5){$g_{\hbox{-}}$}
\put(-135,30.5){$\Omega_{\hbox{-}}$}
\put(-113,53.5){$g_{+}$}
\put(-17,30.5){$\Omega_{+}$}
\put(-99,35.5){$\Omega_{c_{\hbox{-}}}$}
\put(-52,35.5){$\Omega_{c_{+}}$}
\put(-82,20.5){$\Omega_{d}$}
\put(-62,20.5){g}
\caption{(Color online) Schematic representation of the atomic medium required for both storage and encoding of photonic qubits. The quantized (classical) fields are represented by full (dashed) arrows. The two EIT-processes are in red, the tripod-system is in black. All the couplings are resonant except the cavity (represented by a full blue arrow) which is far detuned from resonance. For our model calculations, the system corresponds to the atomic states of neon: $|b\> = (2p^5 3s)^3P_0$, $|a_{\hbox{-}}\rangle,$ $|e\rangle$ and $|a_{+}\rangle$ are associated with the triple degenerate states $(2p^5 3p)$ $^3P_1(m\hbox{=}\hbox{-}1,0,\hbox{+}1)$ of life times of 18ns. The ancillary states $|c_{\hbox{-}}\rangle,$ $|d\rangle$ and $|c_{+}\rangle$ correspond with the metastable states $(3p^5 3s)$ $^3P_2(m\hbox{=}\hbox{-}1, 0, \hbox{+}1)$ of life times up to 0.8s. An intense non-resonant $\sigma_{\pm}$ polarized field ($I_{\hbox{\scriptsize{stark}}} \sim 10^8$ W/cm$^{2}$) with a smaller frequency $(\lambda_{\hbox{\scriptsize{stark}}}=612.520$ nm) than the Bohr frequency of the pair $\{ ^3P_0$-$^3P_1\}$ shifts the states such that each field of appropriate polarization drives a unique transition.
} \label{system_universel}
\end{figure}

In one medium, the initial state of the atomic ensemble plus the single photon is
 \begin{equation}
\label{initial_state}
\Psi_0= \alpha |\underline{b},1_{\hbox{-}},0_{+}\rangle+\beta |\underline{b},0_{\hbox{-}},1_{+}\rangle,
\end{equation}
with $|\underline{b}\rangle\hbox{:=}|b^{(1)},\hdots,b^{(N)}\rangle.$ We use a set of resonant laser fields in the configuration shown in Fig. \ref{system_universel}. To simplify the notation, we define the collective states
$
|\underline{s}\rangle\hbox{:=} \sum_{j=1}^N |b^{(1)},\hdots,s^{(j)},\hdots,b^{(N)}\rangle/{\sqrt{N}},
$ $s$ being the labels of excited states $a_{\pm},e$ and of metastable states $c_{\pm}, d.$ The Fock states associated to zero photon are not written except when this is necessary.
The rotating wave Hamiltonian of the system in the subspace spanned by $\{|\underline{b},1_{\pm}\rangle,|\underline{s}\rangle\}$ is written in the interaction picture as
\begin{eqnarray}
\label{hamiltonian_rotation}
& & H\hbox{=} \nonumber \sum_{\ell=+,\hbox{-}}\big[\Omega_{\ell}e^{\hbox{-}i\varphi_{\ell}} |\underline{a}_{\ell}\rangle\langle \underline{c}_{\ell}| + g_{\ell}\sqrt{N} |\underline{a}_{\ell}\rangle\langle \underline{b},1_{\ell}| \big] \\
& & \hbox{+} \sum_{k=c_{+},c_{\hbox{-}},d} \Omega_{k} e^{i\varphi_{k}}|{\underline{k}}\rangle\langle \underline{e}|+h.c.
\end{eqnarray}
The couplings between the medium and the classical laser fields are given by the time dependent Rabi frequencies $\Omega_{k,\ell}$ whereas the couplings with the photonic qubits are quantified by the time independent Rabi frequencies $g_{\ell}.$ The phase of the laser fields is denoted by $\varphi_{k,\ell}.$ The coupling with the cavity is omitted since its frequency is far from resonance. One can see the advantage of using a collective excitation: The corresponding coupling strength exceeds that of an individual atom by the square root of the number of atoms.
The adiabatic evolution of the initial state (\ref{initial_state}) is completely described by the dark states of the Hamiltonian (\ref{hamiltonian_rotation}), labelled $\Phi_{d}^{(q)}(t)$ (with q the index of the degeneracy). These dark states are instantaneous eigenstates that don't have any components on the atomic excited states. They are associated to null eigenvalues. Although these dark states can be degenerate, they here evolve without any geometric phase, since
$
\langle\phi_d^{(q)}(s)|\frac{d}{ds}|\phi_d^{(q)}(s)\rangle=0.
$
Furthermore, they do not acquire non-abelian holonomy, $\langle\phi_d^{(q')}(s)|\frac{d}{ds}|\phi_d^{(q)}(s)\rangle=0$ since for $q \neq q',$ the dark states belong to orthogonal subspaces. Therefore, according to the adiabatic theorem, the dynamics follows the dark state initially connected to each component of the initial state vector (\ref{initial_state}) 
\begin{equation}
\label{state_vector}
\Psi(t) \approx \sum_{d,q}c_d^{q} \Phi_d^{(q)}(t),
\end{equation}
with $c_{d}^{q} =\<\Phi_d^{(q)}(t_i)|\Psi(t_i)\>.$\\
\subsection{Storage} Each circular polarization components can be stored in a specific atomic collective state \cite{michael_prl2000&opt_comm00}. The storage process is based on EIT-technique in which a medium driven by a control field is rendered transparent for a given frequency. Here, two control fields associated with the Rabi frequencies $\Omega_{\pm}$ are switched off such that 
$
\theta := \arctan g_{\pm}\sqrt{N}/\Omega_{\pm}
$
goes from 0 to $\pi/2$ (see Fig. \ref{simulation_rotation}). Under the adiabatic condition $|\dot{\theta}| \ll (g^2_{\pm}N+\Omega_{\pm}^2)^{1/2},$ the dynamics associated to each components of $\Psi_0$ follows a dark state of the form
$
\cos \theta |\underline{b},1_{\pm}\rangle - \sin \theta e^{i\varphi_{\pm}} |\underline{c}_{\pm}\rangle
$
and the state vector reads
\begin{equation}
\label{Psi_1}
\Psi_1=-\left({\alpha} e^{i\varphi_{\hbox{-}}} |\underline{c}_{-}\rangle+{\beta} e^{i\varphi_{+}} |\underline{c}_{+}\rangle\right).
\end{equation}
The photonic excitation is thus stored in the atomic medium. It can be released by switching on the control fields (with the phase $\varphi_{\pm}'$) such that $\theta$ goes from $\pi/2$ to 0. If the phases of $\Omega_{\pm}$ are unchanged $(\varphi_{\pm}\hbox{=}\varphi_{\pm}'),$ the system returns back into its initial state (\ref{initial_state}). The condition $g_{\pm}\sqrt{N}, \Omega_{\pm}^{\max} \gg 1/\tau,$ $\tau$ being the life time of the excited states $|a_{\pm}\>,$ required to insure the coherence of the storage process can be fulfilled for dense enough medium and strong enough field intensities. \\
\subsection{Encoding} When the photonic excitation is stored, additional laser pulses in a tripod-type configuration \cite{unanyan_opt_comm_98} can be used to manipulate the storage medium in order to prepare the qubit in any preselected superpositon states. The procedure \cite{kis_pra02} consists of two stimulated Raman adiabatic passage (STIRAP) \cite{bergmann1_2}. In the first one, the pulses $\Omega_{c_{\pm}}$ evolving in time with a constant ratio, follow the pulse $\Omega_d.$ Under the adiabatic condition
$|\dot{\chi}| \ll (\Omega^2_{c_{\hbox{-}}}+\Omega^2_{d}+\Omega^2_{c_{+}})^{1/2},$ the dynamics follows two dark states of the form
$
e^{i\varphi_{c_{\hbox{-}}}} \cos \xi \sin \chi |\underline{c}_{-}\rangle+e^{i\varphi_{c_{+}}} \sin \xi \sin \chi |\underline{c}_{+}\rangle-e^{i\varphi_{d}} \cos \chi |\underline{d}\rangle 
$
and 
$
-e^{i\varphi_{c_{\hbox{-}}}} \sin \xi |\underline{c}_{-}\rangle+e^{i\varphi_{c_{+}}} \cos \xi|\underline{c}_{+}\rangle
$
where $\tan\chi:=\Omega_d/(\Omega_{c_{\hbox{-}}}^2+\Omega_{c_{+}}^2)^{1/2}$ and $\tan{\xi}:= \Omega_{c_{+}}/{\Omega_{c_{\hbox{-}}}}.$ At the end of the sequence, the system is in a superposition of the three metastable states $|\underline{c}_{\pm}\rangle,$ $|\underline{d}\rangle.$ The weight and the phase of each components depend on the ratio between $\Omega_{c_+}$ and $\Omega_{c_{\hbox{-}}}$ and on the relative phase of pulses. The second step is a reverse STIRAP process which transfers back the population from $|\underline{d}\rangle$ to $|\underline{c}_{\pm}\rangle.$ The pulses $\Omega_{c_\pm}$ are switched on first and the pulse $\Omega_d$ partially overlaps them. The ratio of the Rabi frequencies $\Omega_{c_\pm}$ is unchanged and the dynamics follows the same dark states that previously. The phase of $\Omega_d$ is changed from $\varphi_d$ in the first STIRAP to $\varphi_d'$ in the second one whereas the phase of $\Omega_{c_\pm}$ is unchanged $(\varphi_{c_\pm}\hbox{=}\varphi_{c_\pm}')$. By transferring back the population from $|\underline{c}_{\pm}\rangle$ to $|\underline{b},1_{\pm}\rangle$ (such that $\varphi_{\pm}=\varphi_{\pm}'),$ the state vector (\ref{Psi_1}) reads
\begin{eqnarray}
\label{Psi_2}
& &\Psi_2 = \nonumber e^{\hbox{-}i/2} \Big(\big[\alpha \left(\cos(\Delta/2)-i\sin(\Delta/2)\cos 2\xi \right) \\ \nonumber
& & -i \beta \sin(\Delta/2) \sin 2\xi e^{-i\eta}\big] |\underline{b},1_{\hbox{-}}\rangle+
\big[-i \alpha \sin(\Delta/2) \sin 2\xi e^{i\eta} \\
& &+ \beta \left(\cos(\Delta/2)+i \sin(\Delta/2) \cos 2\xi\right)\big] |\underline{b},1_{+}\rangle\Big)
\end{eqnarray} 
where $\Delta\hbox{:=}\varphi_{d}'-\varphi_{d},$ $\eta\hbox{:=}\varphi_{c_{+}} - \varphi_{c_{\hbox{-}}}.$ 
This state is thus of the form
$
\Psi_2=e^{\hbox{-}i \Delta/2} U(\Delta, \textbf{n}) \Psi_0,
$
with 
$
U(\Delta,\textbf{n})=e^{-i\frac{\Delta}{2} \textbf{n.}{\bf{\hat{\sigma}}}}.
$
\textbf{n} is the vector $\textbf{n}=(\sin{2\xi} \cos{\eta}, \sin{2\xi} \sin{\eta}, \cos{2\xi})$ and $\hat{\sigma}$ denotes Pauli's spin matrix $\hat{\sigma}=(\sigma_{x}, \sigma_{y}, \sigma_{z}).$ We get a technique to rotate a generic initial state of angle $\Delta$ around the vector \textbf{n}. $U$ constitutes a general transformation of SU(2) from which all one-qubit gates can be obtained by controlling the relative phase of the fields and their polarizations.\\ 
\section{Computation} 
\begin{figure}[ht!]
{\includegraphics[scale=0.35]{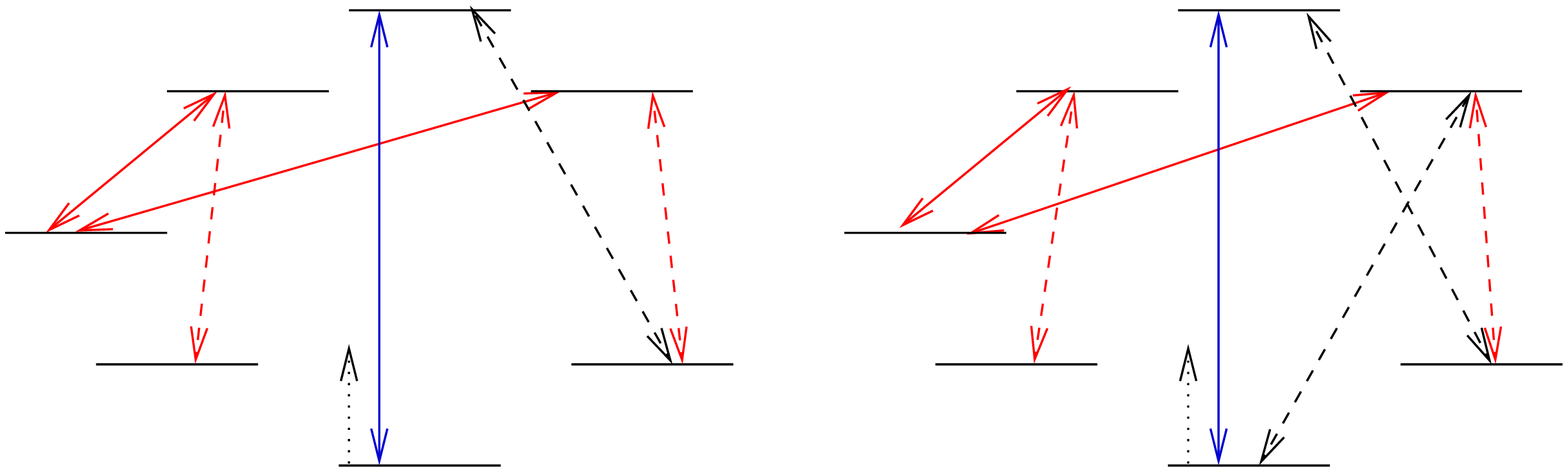}} 
\put(-237,27.5){$|b\rangle$}
\put(-216.5,65.5){$|a_{\hbox{-}}\rangle$}
\put(-158.5,65.5){$|a_{+}\rangle$}
\put(-187.5,79.5){$|e\rangle$}
\put(-186.5,-9.5){$|d\rangle$}
\put(-222.5,6.5){$|c_{\hbox{-}}\rangle$}
\put(-148.5,6.5){$|c_+\rangle$}
\put(-231,50.5){$g_{\hbox{-}}$}
\put(-211,25.5){$\Omega_{\hbox{-}}$}
\put(-203,50.5){$g_{+}$}
\put(-136,25.5){$\Omega_{+}$}
\put(-172,44.5){$\Omega_{c_{+}}$}
\put(-198,5.5){S}
\put(-184,20.5){g}
\put(-205,-25.5){medium 1}
\put(-106,27.5){$|b\rangle$}
\put(-82.5,65.5){$|a_{\hbox{-}}\rangle$}
\put(-27.5,65.5){$|a_{+}\rangle$}
\put(-55.5,79.5){$|e\rangle$}
\put(-55.5,-9.5){$|d\rangle$}
\put(-92.5,6.5){$|c_{\hbox{-}}\rangle$}
\put(-17.5,6.5){$|c_+\rangle$}
\put(-98,50.5){$g_{\hbox{-}}$}
\put(-80,25.5){$\Omega_{\hbox{-}}$}
\put(-68,50.5){$g_{+}$}
\put(-10,25.5){$\Omega_{+}$}
\put(-43,44.5){$\Omega_{c_{+}}$}
\put(-38,8.5){$\Omega_{a_{+}}$}
\put(-67,5.5){S}
\put(-52,20.5){g}
\put(-65,-22.5){medium 2}
\caption{(Color online) Schematic representation of two atomic medium coupled by a cavity. An additional Stark
radiation pulse (represented by a dotted arrow)
introduces a Stark shift of the state $|d\>$ by non-resonant
couplings to all other states outside the seven-level system in such a way that the cavity becomes resonant. We neglect the shifts of the other states induced by the Stark laser. Experimentally this negligence is justified, as the frequency of
the Stark laser can be appropriately chosen. All the other couplings are resonant.} \label{system_cphase}
\end{figure}

We give in this section a sequence of pulses offering the possibility to implement a Cphase gate. This gate associated with the previous one-qubit gates define a universal set \cite{explanation}. \\
\\The generic state of two photons interacting with two spatially separated media is written as
\begin{eqnarray}
\label{initial_2qubit_state}
& & \psi_0=|\underline{b}^{(1)},\underline{b}^{(2)}\> \otimes \\ & &\nonumber
\big[\alpha |1_{\hbox{-}}^{(1)},1_{\hbox{-}}^{(2)}\>+ \beta |1_{\hbox{-}}^{(1)}1_{+}^{(2)}\>+ \gamma |1_{+}^{(1)},1_{\hbox{-}}^{(2)}\>+\delta |1_{+}^{(1)},1_{+}^{(2)}\>\big].
\end{eqnarray}
The superscript (p), p=1,2 labels the medium. We use a set of resonant laser fields in the configuration shown in Fig. \ref{system_cphase}. 
An additional radiation pulse introduces a Stark shift of the state $|d\>$ by non-resonant
couplings to all other states outside the seven-level system in such a way that the cavity mode (associated to the Rabi frequency $\hbox{g}^{(p)},$ p=1,2) becomes resonant. The rotating wave Hamiltonian is written in the interaction picture as
\begin{widetext}
\begin{eqnarray}
\label{hamiltonian_cphase}
&&\nonumber H\hbox{=}
\sum_{p=1,2} \big[ \hbox{g}^{(p)} a_{\hbox{\scriptsize{cav}}}|\underline{e}^{(p)}\rangle\langle \underline{d}^{(p)}|+ \Omega_{\underline{c}_{+}}^{(p)} e^{i\varphi_{c_{+}}^{(p)}}|\underline{c}_{+}^{(p)}\rangle\langle \underline{e}^{(p)}| \hbox{+} \\
&&\sum_{\ell=\pm}\big(\Omega_{\ell}^{(p)}e^{\hbox{-}i\varphi_{\ell}^{(p)}} |\underline{a}_{\ell}^{(p)}\rangle\langle \underline{c}_{\ell}^{(p)}| \hbox{+} g_{\ell}^{(p)} \sqrt{N^{(p)}} |\underline{a}_{\ell}^{(p)}\rangle\langle \underline{b}^{(p)},1_{\ell}^{(p)}| \big)\big] \hbox{+} \Omega_{a_{+}}^{(2)} e^{-i\varphi_{a_{+}}^{(2)}}|\underline{a}_{+}^{(2)}\rangle\langle \underline{d}^{(2)}|\hbox{+h.c.} 
\end{eqnarray}
\end{widetext}
where $a_{\hbox{\scriptsize{cav}}}$ is the annihilation operator
for the cavity mode. The associated Fock states labelled by $|f_{\hbox{\scriptsize{cav}}}\>$ ($f$ corresponding to the number of photons in the cavity mode) are not written except when this is necessary.
The construction of the Cphase gate consists of five steps. For each step, we give the explicit expression of dark states involved in the dynamics. The state vector is deduced from (\ref{state_vector}).\\
$\bullet$ First, we apply the storage technique on both media. The state (\ref{initial_2qubit_state}) reads
\begin{eqnarray}
\label{psi_1_cphase}
& & \nonumber \psi_1= \alpha e^{i\left(\varphi_{\hbox{-}}^{(1)}+\varphi_{\hbox{-}}^{(2)}\right)} |\underline{c}_{{-}}^{(1)},\underline{c}_{-}^{(2)}\rangle+\\ \nonumber
& &\beta e^{i\left(\varphi_{\hbox{-}}^{(1)}+\varphi_{+}^{(2)}\right)} |\underline{c}_{-}^{(1)},\underline{c}_{+}^{(2)}\rangle+\gamma e^{i\left(\varphi_{+}^{(1)}+\varphi_{\hbox{-}}^{(2)}\right)}\times \\
& & |\underline{c}_{+}^{(1)},\underline{c}_{-}^{(2)}\rangle+\delta e^{i\left(\varphi_{+}^{(1)}+\varphi_{+}^{(2)}\right)} |\underline{c}_{+}^{(1)},\underline{c}_{+}^{(2)}\rangle.
\end{eqnarray}
$\bullet$ The population of state $|\underline{c}_{+}^{(2)}\rangle$ (medium 2) is transferred into $|\underline{d}^{(2)}\rangle$ by a STIRAP process generated by the sequence $\Omega_{a+}$-$\Omega_+$ in this order. If $((\Omega_{+}^{(2)\max})^2+(\Omega_{a_{+}}^{(2)\max})^2)^{1/2} \gg g_{+}^{(2)}\sqrt{N},$ $|\underline{b}^{(2)},1_{+}^{(2)}\>$ is not coupled to the other states. Under the adiabatic condition $|\dot\rho| \ll ((\Omega_{+}^{(2)})^2+(\Omega_{a_{+}}^{(2)})^2+(g_{+}^{(2)})^2N)^{1/2},$ $\tan\rho:=\Omega_{+}^{(2)}/\Omega_{a_{+}}^{(2)}$ the dynamics of states involving $|\underline{c}_{+}^{(2)}\rangle$ follows a dark state of the form
$
e^{i\varphi_{+}^{(2)}} \cos\rho |\underline{c}_{+}^{(2)}\rangle-e^{i\varphi_{a_{+}}^{(2)}} \sin \rho |\underline{d}^{(2)}\rangle.
$
The relative phase of the two pulses $\Omega_{a_{+}}$-$\Omega_{+}$ is taken equal to $\pi$ such that the state (\ref{psi_1_cphase}) reads
\begin{eqnarray}
\label{psi_2_cphase}
& & \nonumber \psi_2= \alpha e^{i\left(\varphi_{\hbox{-}}^{(1)}+\varphi_{\hbox{-}}^{(2)}\right)} |\underline{c}_{-}^{(1)},\underline{c}_{-}^{(2)}\rangle \\ \nonumber
& &+\beta  e^{i\left(\varphi_{\hbox{-}}^{(1)}+\varphi_{+}^{(2)}\right)} |\underline{c}_{-}^{(1)},\underline{d}^{(2)}\rangle+\gamma e^{i\left(\varphi_{+}^{(1)}+\varphi_{\hbox{-}}^{(2)}\right)}\times \\ 
& &|\underline{c}_{+}^{(1)},\underline{c}_{-}^{(2)}\rangle+\delta e^{i\left(\varphi_{+}^{(1)}+\varphi_{+}^{(2)}\right)} |\underline{c}_{+}^{(1)},\underline{d}^{(2)}\rangle.
\end{eqnarray}
$\bullet$ The third step adds \textit{a controllable additional phase $\zeta$} to the state $ |\underline{c}_{+}^{(1)},\underline{d}^{(2)}\rangle.$ $\zeta$ define the phase of the $\hbox{Cphase}(\zeta)$ gate. We first transfer the population of $|\underline{c}_{+}^{(1)},\underline{d}^{(2)}\rangle$ into the state $|\underline{d}^{(1)},\underline{c}_{+}^{(2)}\rangle$ by using the pulse sequence $\Omega_{c_{+}}^{(2)}$-$\Omega_{c_{+}}^{(1)}$ in this order. The dynamics associated to each components of state (\ref{psi_2_cphase}) follows a dark state \cite{goto_pra04,sangouard_arxiv}: $|\underline{c}_{-}^{(1)},\underline{c}_{-}^{(2)}\rangle$ and $|\underline{c}_{-}^{(1)},\underline{d}^{(2)}\rangle$ are decoupled from the other states. The dark state $$\Omega_{c_{+}}^{(1)}e^{-i\varphi_{c_{+}}^{(1)}} |\underline{d}^{(1)},\underline{c}_{-}^{(2)}\rangle|1_{\hbox{\scriptsize{cav}}}\>-\hbox{g}^{(1)} |\underline{c}_{+}^{(1)},\underline{c}_{-}^{(2)}\rangle|0_{\hbox{\scriptsize{cav}}}\>$$ is connected to the state $|\underline{c}_{+}^{(1)},\underline{c}_{-}^{(2)}\rangle$ at the beginning and at the end of the sequence. As a consequence, the phase and the weight of $|\underline{c}_{-}^{(1)},\underline{c}_{-}^{(2)}\rangle,$ $|\underline{c}_{-}^{(1)},\underline{d}^{(2)}\rangle$ and $|\underline{c}_{+}^{(1)},\underline{c}_{-}^{(2)}\rangle$ are unchanged. The population transfer from $|\underline{c}_{+}^{(1)},\underline{d}^{(2)}\rangle$ to $|\underline{d}^{(1)},\underline{c}_{+}^{(2)}\rangle$ is realized via the dark state \cite{pellizzari_prl95} 
\begin{eqnarray}
&&\nonumber \hbox{g}^{(1)}\Omega_{c_{+}}^{(2)}e^{-i\varphi_{c_{+}}^{(2)}} |\underline{c}_{+}^{(1)},\underline{d}^{(2)}\rangle|0_{\hbox{\scriptsize{cav}}}\>+\\
&&\nonumber \hbox{g}^{(2)} \Omega_{c_{+}}^{(1)}e^{-i\varphi_{c_{+}}^{(1)}} |\underline{d}^{(1)},c_{+}^{(2)}\rangle|0_{\hbox{\scriptsize{cav}}}\>-\\
&&\nonumber \Omega_{c_{+}}^{(1)}\Omega_{c_{+}}^{(2)} e^{-i(\varphi_{c_{+}}^{(1)}+\varphi_{c_{+}}^{(2)})} |\underline{d}^{(1)},\underline{d}^{(2)}\rangle|1_{\hbox{\scriptsize{cav}}}\>.
\end{eqnarray}
Then, we transfer back the population from state $|\underline{d}^{(1)},\underline{c}_{+}^{(2)}\rangle$ to $|\underline{c}_{+}^{(1)},\underline{d}^{(2)}\rangle$
by using the same pulses in the reverse order $\Omega_{c_{+}}^{(1)}$-$\Omega_{c_{+}}^{(2)}.$ 
The phase of the pulses is changed from $\varphi_{c_{+}}^{(1)},\varphi_{c_{+}}^{(2)}$ in the first sequence to $\varphi_{c_{+}}^{(1)'},\varphi_{c_{+}}^{(2)'}$ in the second one. We define $\zeta:=\varphi_{c_{+}}^{(2)}-\varphi_{c_{+}}^{(2)'}-\varphi_{c_{+}}^{(1)}+\varphi_{c_{+}}^{(1)'}.$ The dynamics follows the same dark states that previoulsy. At the end of the sequence, the state vector $\psi_3$ is defined by (\ref{psi_2_cphase}) where $|\underline{c}_{+}^{(1)},\underline{d}^{(2)}\rangle$ has to be replaced by $e^{i\zeta}|\underline{c}_{+}^{(1)},\underline{d}^{(2)}\rangle.$ The strong coupling regime $\hbox{g}^{(p)}, \Omega_{c_{+}}^{(p)\max} \gg 1/\tau, \kappa$ (where $\kappa$ is the decay rate of the cavity field and $\tau$ is here the life time of $|e\>$), has to be fulfilled to insure the coherence of the process. Recent technological advances in optical cavity QED allow one to achieve this regime \cite{mckeever_Nat03} making the process doable in practise. Furthermore, if $\hbox{g}^{(p)} \gg \Omega_{c_{+}}^{(p)\max},$ the cavity is negligibly populated and the coupling between the media is provided by a virtual photon. \\
$\bullet$ The population of $|\underline{d}^{(2)}\rangle$ (medium 2) is transferred back to $|\underline{c}_{+}^{(2)}\rangle.$ The sequence of pulses required is $\Omega_{+}^{(2)}$-$\Omega_{a_{+}}^{(2)}$ chosen with a relative phase equal to $\pi.$ The state vector is defined as (\ref{psi_1_cphase}) by replacing $|\underline{c}_{+}^{(1)},\underline{c}_{+}^{(2)}\rangle$ by $e^{i\zeta}|\underline{c}_{+}^{(1)},\underline{c}_{+}^{(2)}\rangle.$ \\
$\bullet$ The two photons are finally released. If the phase of the control fields $\Omega_{\pm}^{(p)}$ is unchanged $(\varphi_{\pm}^{(p)}\hbox{=}\varphi_{\pm}^{(p)'},$ p=1,2), the state vector
\begin{eqnarray}
& & \psi_5=|\underline{b}^{(1)},\underline{b}^{(2)}\> \otimes \\ & &\nonumber
\big[\alpha |1_{\hbox{-}}^{(1)},1_{\hbox{-}}^{(2)}\>+ \beta |1_{\hbox{-}}^{(1)}1_{+}^{(2)}\>+ \gamma |1_{+}^{(1)},1_{\hbox{-}}^{(2)}\>+\delta e^{i\zeta}|1_{+}^{(1)},1_{+}^{(2)}\>\big]
\end{eqnarray}
coincides with the output state of a $\hbox{Cphase}(\zeta)$ gate.
\section{Numerical simulation}
\begin{figure}[ht!]
{\includegraphics[scale=0.5]{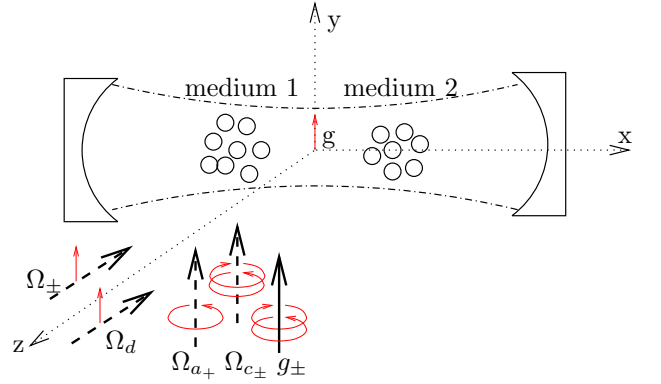}} 
\put(-135,-5.5){$g_{\pm}$}
\put(-155,-5.5){$\Omega_{c_{\pm}}$}
\put(-175,-5.5){$\Omega_{a_{+}}$}
\put(-230,25.5){$\Omega_{\pm}$}
\put(-200,2.5){$\Omega_{d}$}
\put(-118,80.5){g}
\put(-6,80.5){x}
\put(-116,124.5){y}
\put(-235,1.5){z}
\put(-110,98.5){medium 2}
\put(-170,98.5){medium 1}
\caption{(Color online) Geometry of the experimental setup and polarizations of the fields. The circles represent the atoms trapped in the optical cavity. The (black) thick arrows give the propagation direction whereas the (red) thin arrows are associated with the polarizations of the fields.} \label{geometry_setup}
\end{figure}

Neon atom is a potential candidate for the experimental realization of the proposed protocol. The labels of the states are given in the caption of Fig. \ref{system_universel}. The spectroscopic values of the considered transitions can be found in ref. \cite{Heinz_thesis}. The polarizations of the laser fields and of the cavity are compatible with the geometry of the experimental setup. Indeed, we define a x-axis parallel to the optical axis of the cavity and a y-axis corresponding to a $\pi$-polarization (see Fig. \ref{geometry_setup}). The $\pi$-polarized fields $(\Omega_\pm, \Omega_d)$ can propagate along the z-axis whereas the $\sigma_\pm$-polarized fields $(g_{\pm}, \Omega_{c_{\pm}},\Omega_{a_{+}})$ can propagate along the y-axis.

\begin{figure}[ht!]
{\includegraphics[scale=0.5]{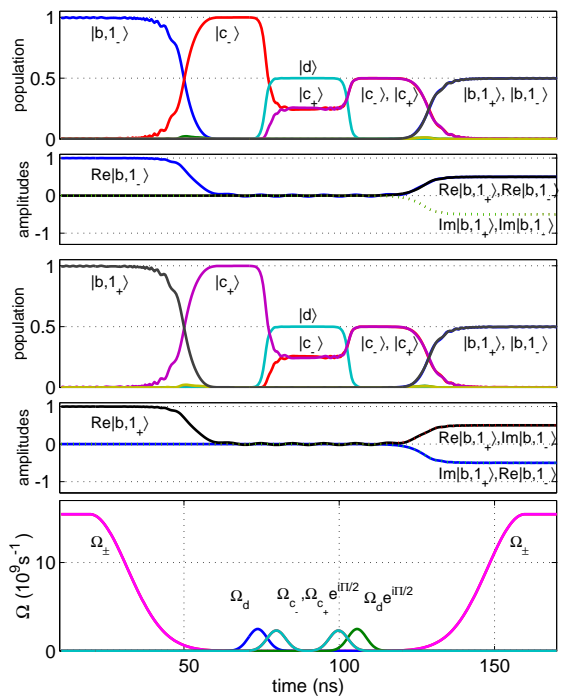}} 
\caption{(Color online) Numerical simulation of a $\pi/4$-rotation gate. From top to bottom: Temporal evolution of population and amplitudes of initial states $|\underline{b},1_{\hbox{-}}\rangle$ and $|\underline{b},1_{+}\rangle.$ The sequence of pulses is given in the last frame. The pulses have Gaussian shapes of FWHM and intensities $T_{\pm}=20$ns and $3.24\times10^4$W/cm$^{2}$ for $\Omega_{\pm}$ ; $T=5$ns and $7.29\times10^2$W/cm$^{2}$ for $\Omega_{c_{\pm}}$ and $\Omega_d$. These parameters have been chosen to fulfil the adiabatic conditions and the EIT requirement: $g_{\pm}\sqrt{N} T_{\pm}= 12,$ $\Omega_{\pm}^{\max}/g_{\pm}\sqrt{N}=25,$ $\Omega_{c_{\pm}}^{\max} T=12,$ $\Omega_d^{\max} T=12.$ Since the pulses $\Omega_{c_{\hbox{-}}}$ and $\Omega_{c_{+}} $ are used two times successively in a constant ratio, they could be replace by a single pulse. The process would require the use of only five pulses.} \label{simulation_rotation}
\end{figure}

We present in Fig. \ref{simulation_rotation}, the numerical validation of a $\pi/4$-rotation gate. The couplings $\Omega_{\pm}$ $(\Omega_{c_{\pm}})$ are induced by a unique field with a $\pi$$(\sigma_\pm)$-polarization. The construction of the desired gate requires the use of the following pulse parameters: (i) $\varphi_{\pm}=\varphi_{\pm}',$ (ii) $\Omega_{c_{+}}/\Omega_{c_{\hbox{-}}} = 1$ such that $2\xi= \pi/2$ and (iii) $\varphi_{c_{+}}=\varphi_{c_{\hbox{-}}}+\pi/2$ such that $\eta=\pi/2.$ 
Up to an irrelevant phase, a rotation gate of angle $\Delta/2$ is obtained. The rotation angle is thus controlled by the relative phase $\varphi_{d}'-\varphi_{d}$ here chosen equal to $\pi/2.$\\

\section{Conclusion}
In summarize, we have presented a scheme based on adiabatic passage along dark states to encode the qubits, to store them and to compute an arbitrary quantum calculation. 
The universal set of gates required for the encoding and the computation has been implemented in a robust manner by controlling precisely the parameters that determine the action of the gates. Therefore, we did not use dynamical phases, requiring controllable field amplitudes, nor
geometrical phases, requiring a controllable loop in the parameter
space \cite{Unanyan1,Duan,Unanyan2, Pachos}. We used instead static
phase differences of lasers, which can be easily controlled
experimentally.\\
The proposed  protocol could be scaled up to many atoms by placing parts of the register in different cavities and by entangling two storage media of two separated cavities. This protocol offers a promising, robust and decoherence-free approach for the realization of any quantum circuits.
\begin{acknowledgments}
I am indebted to Z. Kis and M. Fleischhauer for enlighting discussions and criticisms of the manuscript. I acknowledge X. Lacour, M. Heinz, F. Vewinger and R. Unanyan for valuable advices and supports from the EU network QUACS under contract No. HPRN-CT-2002-0039 and from La Fondation Carnot.
\end{acknowledgments}

\end{document}